\documentclass [prl,10pt,twocolumn] {revtex4}
\usepackage{amssymb}
\usepackage{amsmath}
\usepackage{subfigure}
\input{epsf}
\usepackage{graphicx}
\begin{document}

\title{Entanglement Spectrum as a Generalization of Entanglement Entropy: Identification \\of Topological Order in Non-Abelian Fractional Quantum Hall Effect States}
\author{Hui Li}
\affiliation{Physics Department, Princeton University, Princeton, New Jersey 08544, USA}
\author{F. D. M. Haldane}
\affiliation{Physics Department, Princeton University, Princeton, New Jersey 08544, USA}
\date{July 3, 2008}
\begin{abstract}
We study the ``entanglement spectrum'' (a presentation of the
Schmidt decomposition analogous to a set of ``energy levels'') of a 
many-body state, and compare the 
Moore-Read model wavefunction for the
$\nu$ = 5/2 fractional quantum Hall state with a generic 5/2 state
obtained by finite-size diagonalization of the second-Landau-level-projected
 Coulomb interactions.
Their spectra  share a common ``gapless'' structure, related to 
conformal field theory.   In the model state, 
these are the \textit{only} levels, while in the ``generic''  case, they
are separated from the 
rest of the spectrum by a clear ``entanglement gap'', 
which appears to remain finite in the thermodynamic limit.
We propose that the low-lying entanglement spectrum can be used as a
``fingerprint'' to identify 
topological order.
\end{abstract}
\maketitle

There has been increasing recent interest in using quantum entanglement as a probe to detect topological properties of many-body quantum states \cite{ref7, ref7-2, ref7-3, ref7-4}, in particular the states exhibiting the fractional quantum Hall effect (FQHE) \cite{ref1,ref1-2,ref2,ref3,ref3-2,ref3-3,ref5,ref6,ref6-2}. 
Among various measures of quantum entanglement, the entanglement entropy has by far been the favorite \cite{ref6,ref6-2,ref6-3}. By partitioning a many-body quantum system into two blocks, the entanglement entropy is defined as the von Neumann entropy of the reduced density-matrix of either one of the two blocks, and is a single number that can be obtained from knowledge of the density-matrix eigenvalue spectrum. The density matrix may be written in the form $\hat{\rho}=\exp(-\hat{H})$, so that the entanglement entropy is equivalent to the thermodynamic entropy of a system described by a hermitian ``Hamiltonian'' $\hat{H}$ at ``temperature'' $T=1$; in the case of a weak entanglement, the ``excited states'' eigenvalues of $\hat{H}$ are separated from the ground state eigenvalue by a large ``energy gap'' that becomes infinite in the limit of a simple product state with vanishing entanglement entropy. In this Letter, we point out that the spectrum of the ``Hamiltonian'' $\hat{H}$, which we call the {\it ``entanglement spectrum''}, reveals much more complete information than the entanglement entropy, a single number.

We examined the entanglement spectrum of the ideal Moore-Read (MR) state
with Landau-level filling fraction $\nu$ = 1/2, and found that its
structure appears to coincide with that of the
spectrum of the associated nonabelian conformal field theory (CFT). 
In particular, the spectrum is ``gapless'', and (up to a limit determined by the compactification of the system to have a finite area) has the same count of states (characters) at each momentum (Virasoro level), although there is 
some splitting of the ``energies''. 
The MR entanglement spectrum  is also ``non-generic'', as it contains far fewer levels than the number expected from counting Hilbert-space dimensions; this is a common property of model wavefunctions, which  lack part of the
``zero-point fluctuation'' of a generic state.

We also examined the ``realistic'' $\nu$ = 5/2 = 2 + 1/2 state 
obtained by diagonalizing 
the Coulomb interaction projected into the half-filled second Landau level, and observed that all the extra levels expected in a generic entanglement spectrum are present, but are separated by a finite gap from ``gapless'' CFT-like low-lying states with essentially the same structure as those of the Moore-Read state. This gap appears to remain finite in the thermodynamic limit. Thus the low-lying entanglement spectrum is essentially a ``fingerprint'' which allows the associated CFT (which characterizes the topological order) to be identified.

If Landau-level mixing is  ignored, the $\nu=5/2$ FQHE system is equivalent 
to a $\nu$ = 1/2 lowest Landau level with interaction pseudo-potentials \cite{ref5} corresponding to a simple  Coulomb interaction projected into the second Landau level (we did not refine this to include the quantum well form-factor, 
which may be important for quantitative modeling). In spherical geometry \cite{ref5}, the number of 
electrons ($N_e$) and the total number of Landau level orbitals ($N_{orb}$) 
are related by $N_{orb} = 2N_e-2$. In second-quantization, a many-body state can be written in the basis of orbital occupations; we used the $L_z$ eigenstate
basis to divide the spherical surface at a line of latitude into two regions,
so the $N_{orb}$ orbitals are partitioned into $N^A_{orb}$ around the north 
pole, and $N^B_{orb}$ around the south pole (this is spatially the sharpest cut
consistent with projection into a Landau level), which is a partition of
the Fock space $\mathcal H$ into two parts $\mathcal H_A \otimes \mathcal H_B$.

A Schmidt decomposition (equivalent to the singular value decomposition of a matrix) of a  many-body state $|\psi\rangle$ gives
\begin{equation}
|\psi\rangle = {\sum}_{i}e^{-\frac{1}{2}\xi_{i}}|
\psi_{A}^{i}\rangle\otimes|\psi_{B}^{i}\rangle \label{eq1}
\end{equation}
where $\exp(-\frac{1}{2}\xi_{i}) \geq 0$, 
$|\psi_{A}^{i}\rangle\in\mathcal{H}_{A}$, $|\psi_{B}^{i}\rangle\in\mathcal{H}_{B}$, and 
$\langle\psi_{A}^{i}|\psi_{A}^{j}\rangle=\langle\psi_{B}^{i}|\psi_{B}^{j}\rangle=\delta_{ij}$, giving 
$\exp{(-\frac{1}{2}\xi_{i})}$ 
as the singular values and $|\psi_{A}^{i}\rangle$ 
and $|\psi_{B}^{i}\rangle$ the singular vectors. 
If the state is normalized, ${\sum}_{i}\exp{(-\xi_{i})}=1$, but it is not
necessary to impose the normalization condition.

\begin{figure*}[t]
\begin{center}
\subfigure{\label{mrfig1}\includegraphics[width=0.66\columnwidth]{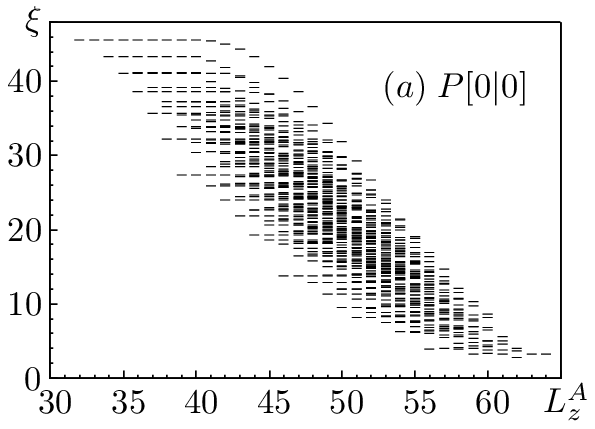}}\hfill
\subfigure{\label{mrfig2}\includegraphics[width=0.66\columnwidth]{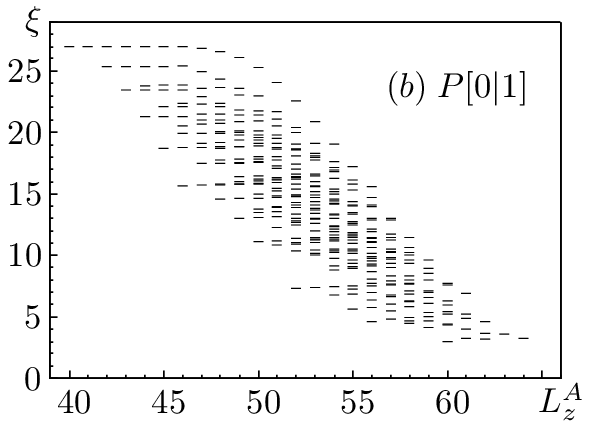}}\hfill
\subfigure{\label{mrfig3}\includegraphics[width=0.66\columnwidth]{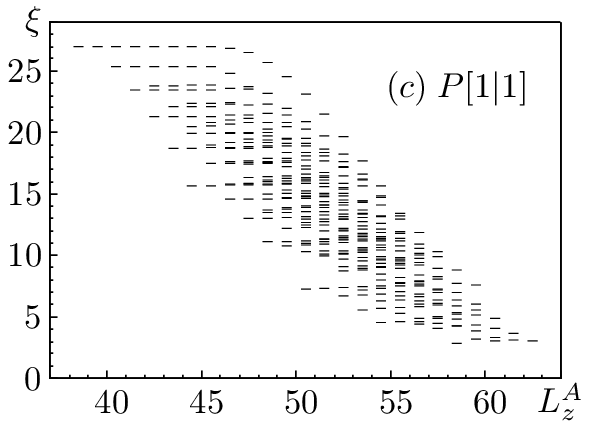}}
\caption{\label{mrfig}The complete entanglement spectra of the $N_e=16$ and $N_{orb}=30$ Moore-Read state (only the relative values of $\xi$ and $L_z^A$ are meaningful).}
\end{center}
\end{figure*}

\begin{table}[b]
\begin{center}
\caption{The numbers in the parenthesis are values of $(N_{orb}^A, N_e^A)$, respectively for each system and partitioning as specified.}
\begin{ruledtabular}
\begin{tabular}{llll}
$N_e$       & $P[0|0]$ & $P[0|1]$ & $P[1|1]$ \\ \hline
$10$ & $(7,4)$ & $(8,4)$ & $(9,5)$ \\
$12$ or $14$ & $(11,6)$ & $(12,6)$ & $(13,7)$ \\
$16$ & $(15,8)$ & $(16,8)$ & $(17,9)$
\end{tabular}
\end{ruledtabular}
\label{table}
\end{center}
\end{table}

The $\xi_{i}$'s are ``energy levels'' of a system with thermodynamic entropy at 
``temperature'' $T=1$ equivalent to the entanglement entropy, 
$\mathcal{S}={\sum}_{i}\xi_{i}\exp{(-\xi_{i})}$, which has been shown to contain information on the topological properties of the many-body state \cite{ref6}. 
The full  structure of the ``entanglement spectrum'' (logarithmic Schmidt spectrum) of 
levels $\xi_i$ contains much more information than the entanglement entropy $\mathcal{S}$, 
a single number. 
This is analogous to the extra information about a condensed matter system 
given by its low-energy excitation spectrum rather than just by its ground state energy.

Because the FQHE ground state is translationally and rotationally invariant
(with quantum number $L_{tot}$ = 0 on the sphere), and
the partitioning of Landau-level orbitals conserves 
both  gauge symmetry and 
rotational symmetry along the $z$-direction, 
in either block $A$ or $B$ both the electron number ($N_e^A$ and $N_e^B$) 
and the total $z$-angular momentum ($L^{A}_{z}$ and $L^{B}_{z}$) are 
good quantum numbers constrained by
$N^A_{e}$ + $N^B_{e}$ = $N_e$, $L^A_z$ + $L^B_z$ = 0.
The entanglement spectrum splits
into distinct sectors labeled by  $N^A_e$ and $L^A_z$.

In the thermodynamic limit, the MR model state can be represented 
by its ``root configuration'' \cite{ref8}, which has occupation numbers 
``$11001100\cdots 110011$'', with a repeated sequence $\cdots 1100\cdots$;
in spherical geometry, this
is terminated by ``$11$'' at both ends.
This is the highest-density  ``MR root configuration'', which we define
as an occupation-number configuration satisfying a ``generalized Pauli 
principle''
that \textit{no group of 4 consecutive orbitals contains more than 2 particles} (this
 rule also applies to MR states with quasiholes, and generates the CFT edge
 spectrum of a finite MR droplet on the open plane \cite{ref8}.)
When the MR state is expanded in the occupation-number basis,
the only configurations (Slater determinants) present are those obtained by starting from
the root configuration, and ``squeezing'' pairs of particles with $L_z$ = 
$m_1,m_2$ closer together,
reducing $|m_1-m_2|$, while preserving $m_1+m_2$ \cite{ref8}.

\begin{figure*}[t]
\begin{center}
\subfigure{\includegraphics[width=0.66\columnwidth]{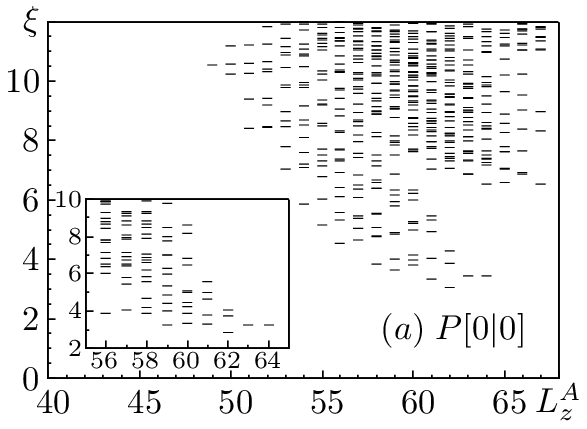}}\hfill
\subfigure{\includegraphics[width=0.66\columnwidth]{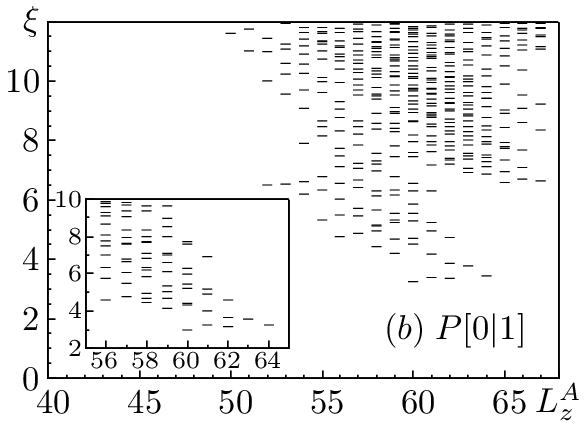}}\hfill
\subfigure{\includegraphics[width=0.66\columnwidth]
{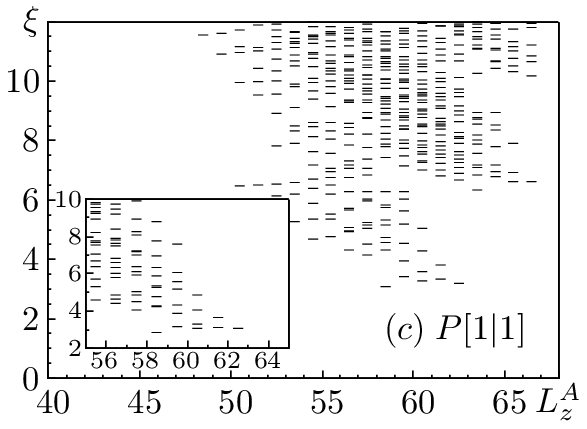}}
\caption{\label{coloumbfig}
The low-lying entanglement spectra of the $N_e=16$ and $N_{orb}=30$ ground state of the Coulomb interaction projected into the second Landau level (there are levels beyond the regions shown here, but they are not of interest to us). 
The insets show the low-lying parts of the spectra of the Moore-Read state, for comparison [see Figure (\ref{mrfig})]. 
Note that the structure of the low-lying spectrum is essentially identical to that of the ideal Moore-Read state.}
\end{center}
\end{figure*}

From the root configuration, 
we see that there are three distinct ways of partitioning the orbitals: (i) between two 0's; (ii) between two 1's; or (iii) between 0 and 1 (the partitioning between 1 and 0 is equivalent to that between 0 and 1 by reflection symmetry). 
We use symbols $P[0|0]$, $P[1|1]$, and $P[0|1]$ to represent the three cases, respectively.  This will correspond to choosing one of the three sectors of the
associated conformal field theory.
For finite systems, we always try to draw the boundary of the partitioning either on the equator (if possible), or closest to the equator but in the southern hemisphere. 
Moreover, we can associate a ``natural'' value to $N_e^A$ for a particular partitioning, i.e., the total number of 1's in the root occupation sequence on the left-hand-side to the boundary. 
In this Letter, it is sufficient to consider only levels whose $N_e^A$ is exactly this natural value. 
Table \ref{table} describes the precise meaning of these symbols for systems that are considered here.

Figure (\ref{mrfig}) shows the spectra for each of the three different ways of partitioning, for the Moore-Read state at $N_e=16$ and $N_{orb} = 30$. 
The spectrum not only has far fewer 
levels than expected for a generic wavefunction, but also exhibits an 
intriguing level-counting structure (as a function of $L^A_z$ and $N^A_e$) 
that resembles that of  
the associated conformal field theory of the edge excitations.
Intuitively, this is  because the boundary of the Landau-level 
partitioning indeed defines an edge shared by region $A$ and $B$. 

In the intuitive picture, the quantum entanglement between $A$ and $B$ arises from {\it correlated quasihole excitations} across the boundary along which the partitioning is carried out. 
Any quasihole excitation in region $A$ necessarily pushes electrons into region $B$, and vice versa. 
However, the electron density 
anywhere on the sphere must remain constant, which can be achieved if the quasihole excitations in $A$ and $B$ are correlated (entangled). 
This gives the empirical rules of counting the levels. 
Take the spectrum in Figure (\ref{mrfig1}) as an example. The partitioning $P[0|0]$ results in the 
root configuration $110011001100110$ on the northern hemisphere (region $A$), and it corresponds to the single ``level'' at the highest possible value of $L_z^A=L_{z,max}^A=64$. 
We measure the $L_{z}^A$ by its deviation from $L_{z,max}^A$, i.e. $\Delta L:=L_{z,max}^A-L_{z}^A$, which has the physical meaning of being the total $z$-angular momentum carried by the quasiholes. 
At $\Delta L=1$, the levels correspond to edge excitations upon the $\Delta L=0$ root configuration. 
There is exactly one edge mode in this case, represented by the MR root configuration $110011001100101$.

The number of $\Delta L=2$ levels can be counted in exactly the same way. 
There are three of them, of which the root configurations are
\begin{eqnarray}
1100110011001001\notag\\
1100110011000110\notag\\
1100110010101010\notag
\end{eqnarray}
while for $\Delta L=3$, the five root configurations are
\begin{eqnarray}
11001100110010001\notag\\
11001100110001010\notag\\
11001100101001100\notag\\
11001100101010010\notag\\
11001010101010100\notag
\end{eqnarray}
The counting for the levels at small $\Delta L$ for $P[0|1]$ and $P[1|1]$ can be obtained similarly.

For an infinite system in the thermodynamic limit, the above idea gives an empirical counting rule of the number of levels at any $\Delta L$, i.e., it is the number of independent quasihole excitations upon the semi-infinite root configuration uniquely defined by the partitioning. 
For a finite system, this rule explains the counting only for small $\Delta L$; for large $\Delta L$, the finite size limits the maximal angular momentum that can be carried by an individual quasihole. 
Therefore the number of levels at large $\Delta L$ in a finite system will be smaller than the number expected in an infinite system. 
Not only is this empirical rule consistent with all our numerical calculation, but it also explains why $P[0|0]$ and $P[0|1]$ have essentially identical low-lying structures. 
This is because the (semi-infinite) configuration ``$\cdots1100110$'' is essentially equivalent to ``$\cdots11001100$'' (with an extra ``$0$'' attached to the right). 
We expect that $P[0|0]$ and $P[0|1]$ become exactly identical in the thermodynamic limit.

For completeness, we list the root configurations associated with the first few low-lying levels in Figure (\ref{mrfig3}).
\begin{eqnarray}
\Delta L=0: \qquad && 11001100110011001 \notag\\
\Delta L=1: \qquad && 110011001100110001 \notag\\
                   && 110011001100101010 \notag\\
\Delta L=2: \qquad && 1100110011001100001 \notag\\
                   && 1100110011001010010 \notag\\
                   && 1100110011001001100 \notag\\
                   && 1100110010101010100 \notag
\end{eqnarray}

Figure (\ref{coloumbfig}) shows the spectra of the system of the same size as in Figure (\ref{mrfig}), i.e., $N_e=16$ and $N_{orb}=30$, but for the ground state of the Coulomb interaction projected into the second Landau level, obtained by direct diagonalization. 
Interestingly, the low-lying levels have the same counting structure as the corresponding Moore-Read case. 
We identify these low-lying levels as the ``CFT'' part of the spectrum, 
in contrast to the other {\it generic}, non-CFT levels that are expected for generic many-body states. 
At relatively small $\Delta L$ (up to a limit which grows with the size of the system), the CFT levels are separated from the generic levels by a clear gap, which we define as the distance from the average of the CFT levels to the bottom of the generic levels. 

\begin{figure*}[t]
\begin{center}
\subfigure{\label{gapfig1}\includegraphics[width=0.66\columnwidth]{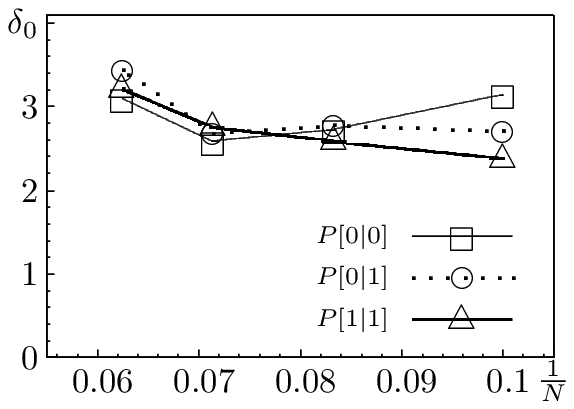}}\hfill
\subfigure{\label{gapfig2}\includegraphics[width=0.66\columnwidth]{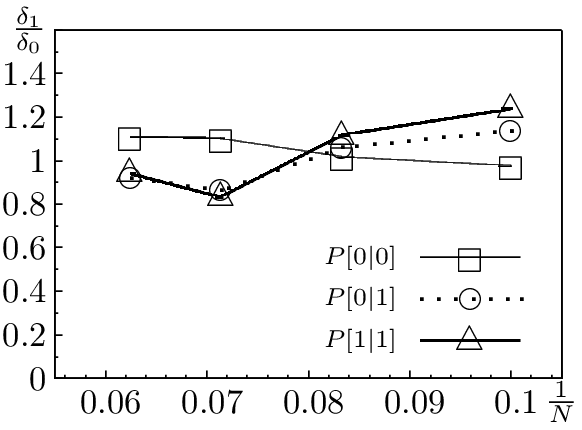}}\hfill
\subfigure{\label{gapfig3}\includegraphics[width=0.66\columnwidth]{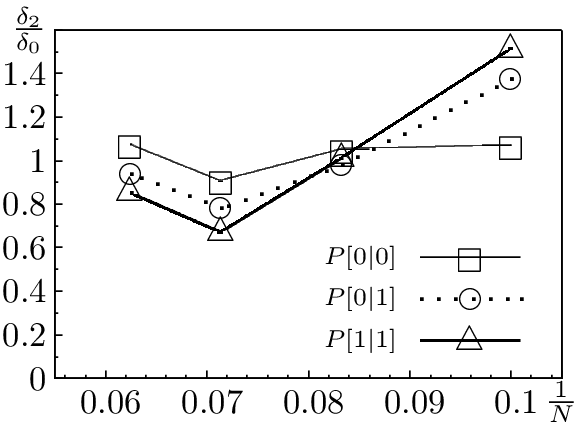}}
\caption{\label{gapfig}Entanglement gap as a function of $1/N$. $\delta_0$ is the gap at $\Delta L=0$, i.e., the distance from the single CFT level at $\Delta L=0$ to the bottom of the generic (non-CFT) levels at $\Delta L=0$. At $\Delta L=1,2$, the gap $\delta_{1,2}$ is defined as the distance from the average of the CFT levels to the bottom of the generic levels. See Table \ref{table} for the details of various partitionings.}
\end{center}
\end{figure*}

Figure (\ref{gapfig}) shows the value of this gap (at $\Delta L=0,1,2$ respectively) as a function of the size of the system, based on which we speculate that the gap between the CFT and non-CFT levels remains finite in the thermodynamic limit for all $\Delta L$. 
The observed fact that the structure of the low-lying spectrum is essentially identical to that of the Moore-Read state, as well as the existence of the ``entanglement gap'', serve as evidence that the ``realistic'' $\nu=5/2$ FQH states is indeed modeled by the Moore-Read state.

Assuming that the gap does remain finite in the thermodynamic limit, characterization of the entanglement spectrum is a reliable way to identify a topologically ordered state. 
While finite-size numerical studies often show impressive (e.g., $99\%$) overlaps between model wave-functions (Laughlin, Moore-Read, etc.) and ``realistic'' states at intermediate system sizes, this cannot persist in the thermodynamic limit. 
Furthermore, the entanglement spectrum is a property of the ground state wave-function itself, as oppose to the physical excitations of a system with boundaries, so allows direct comparison between model states and physical ones.

The asymptotic behavior of the characters of a CFT (the \textit{count} of independent states at each Virasoro level, in each sector) defines both the effective conformal anomaly $\tilde c$ of the CFT, and the quantum dimension of each sector. 
To the extent that there is a clear separation of the ``gapless'', low-lying CFT-like modes and the generic (but ``gapped'') modes, one can count the number of ``gapless'' modes as a function of momentum parallel to the boundary separating the two regions. 
For a finite-size system, these will match the CFT characters up to some limit that grows with system size. 
These numbers are integers, so are not subject to numerical error, and in principle, both $\tilde c$ and the quantum dimensions can be 
extracted from their behavior as the system size grows.

As a critical point is approached, the ``entanglement gap'' may 
still be finite but well below the ``temperature'' $T=1$ at which the 
von Neumann entropy is evaluated. 
We suggest that the direct study of the low-lying entanglement spectrum is a 
far more meaningful way to characterize bipartite entanglement. 
Equivalently, the $T \rightarrow 0$
``low-temperature'' entropy of the modified family of density matrices 
$\hat{\rho}^{(1/T)}$ may prove useful, 
as this  corresponds to the thermodynamic
entropy of the entanglement spectrum at temperature $T$.

We thank B. A.  Bernevig for valuable comments and suggestions. This work was supported in part by NSF MRSEC DMR02-13706.

\end{document}